\documentclass[11pt,twocolumn]{article}      

\usepackage{mathtools, amsfonts}
\usepackage{subcaption}
\usepackage{graphicx}
\usepackage{booktabs}
\usepackage{threeparttable}
\usepackage[switch]{lineno} 
\usepackage[toc]{appendix}
\usepackage{color}
\usepackage[english]{babel}
\usepackage[T1]{fontenc}
\usepackage{amsmath}
\usepackage{subcaption}
\usepackage{amsfonts}
\usepackage{epstopdf}
\usepackage[export]{adjustbox}
\usepackage{amsopn}
\usepackage{float}
\usepackage[a4paper, total={6in, 8in}, margin=0.5in, bottom=0.9in]{geometry}
\usepackage{amssymb}
\usepackage{mathrsfs}

\usepackage{nomencl}

\makenomenclature

\modulolinenumbers[1]

\usepackage{tikz}
\usetikzlibrary{matrix} 
\usetikzlibrary{arrows} 
\usetikzlibrary{calc} 
\tikzstyle{block} = [draw,rectangle,thick,minimum height=2em,minimum width=2em]
\tikzstyle{connector} = [->,thick]

\newcommand{\R}{\mathbb{R}}

\title{Semi-intrusive uncertainty propagation for multiscale models}
\author{Anna Nikishova\footnote{Computational Science Lab, Institute for Informatics, Faculty of Science, University of Amsterdam, The Netherlands, Email:A.Nikishova@uva.nl}  \and Alfons Hoekstra\footnote{Computational Science Lab, Institute for Informatics, Faculty of Science, University of Amsterdam, The Netherlands and ITMO University, Saint Petersburg, Russia}}

\begin{document}

\maketitle

\begin{abstract}
A family of semi-intrusive uncertainty propagation (UP) methods for multiscale models is introduced. The methods are semi-intrusive in the sense that inspection of the model is limited up to the level of the single scale systems, and viewing these single scale components as black-boxes. The goal is to estimate uncertainty in the result of multiscale models at a reduced amount of time as compared to black-box Monte Carlo (MC).
In the resulting semi-intrusive MC method, the required number of samples of an expensive single scale model is minimized in order to reduce the execution time for the overall UP. In the metamodeling approach the expensive model component is replaced completely by a computationally much cheaper surrogate model. 
These semi-intrusive algorithms have been tested on two case studies based on reaction-diffusion dynamics. The results demonstrate that the proposed semi-intrusive methods can result in a significant reduction of the computational time for multiscale UP, while still computing accurately the estimates of uncertainties.
The semi-intrusive methods can therefore be a valid alternative, when uncertainties of a multiscale model cannot be estimated by the black-box MC methods in a feasible amount of time.
\end{abstract}

\section{Introduction}

Computer modeling is widely used in science and engineering to study systems of interest and to predict their behaviour. These systems are usually multiscale in nature, as their accuracy and reliability depend on the correct representation of processes taking place on several length and time scales \cite{Weinan_2011,groen2014survey,hoekstra2014multiscale,karabasov2014multiscale,sloot2009multi}. Moreover, these multiscale systems are usually stochastic, since there are always some unresolved scales, whose effects are not taken into account due to lack of knowledge or limitations of computational power \cite{karabasov2014multiscale,alowayyed2016multiscale}. 
Moreover, measurements of model parameters, model validation, or initial and boundary conditions rarely can be achieved with perfect accuracy \cite{Johnstone_2015}. Therefore, the model results inevitably contain uncertainties, and one should estimate their magnitudes by applying an uncertainty propagation (UP) method. 

Usually a distinction is made between intrusive UP methods, where one substitutes the original model with its stochastic representation, and non-intrusive methods, where the original model is used as a black-box \cite{Le_Ma_tre_2010,smith2013uncertainty}. Intrusive methods are efficient and relatively easy to apply to linear models, e.g. \cite{Wan_2005}. This, however, represents only a relative small class of models. They can be applied to non-linear models as well, but solution of the resulting equations may become very demanding \cite{Xiu09fastnumerical}. Non-intrusive methods can be applied to any type of non-linear model. However, if a single model run requires large execution times, these UP methods may be ineffective, or even computationally intractable. 

In this paper, a family of semi-intrusive UP algorithms for multiscale models is introduced. These methods are called \textit{semi-intrusive}, since they are intrusive only on the level of the multiscale model, that is, in the way the single scale components are coupled together. The single scale components themselves however are treated as black-boxes. 

First, the semi-intrusive Monte Carlo (SIMC) method will be introduced, in which the number of samples for the computationally intensive part of the multiscale model (usually microscale dynamics) will be reduced. This leads to a decrease in the computational time for the multiscale UP. A cross validation, which is part of the method, controls the level of sub-sampling and hence the accuracy of the estimates of uncertainty. 

Next, a metamodeling approach is introduced, where a surrogate model substitutes the most expensive single scale model. The metamodel can be constructed by applying, for example, a data-driven approach, like Gaussian process regression \cite{WANG2015159,Liu2016,Zhan2017}, or using a spectral approach, like the stochastic Galerkin method \cite{GERRITSMA20108333}. Since only one component of the multiscale model is approximated by the surrogate, the resulting error in the model output can be small enough to still be able to obtain reliable uncertainty estimates. However, it is expected that this strongly depends on the sensitivity of the output of the multiscale model on that of the single scale component and the method used to build a surrogate of that single scale component.

These UP methods have been tested on two case studies based on reaction-diffusion dynamics with random inputs: a one-dimensional system with slow diffusion and fast reaction and the two-dimensional Gray-Scott model.

\section{Multiscale model}
\label{sec:MSmodeling} 

According to the Multiscale Modelling and Simulation Framework (MMSF) \cite{Chopard_2011,Borgdorff_2013,BORGDORFF2014719,borgdorff2014performance}, multiscale models can be seen as collections of single scale components coupled through the spatio-temporal scales using scale bridging methods. In the current work, this representation of multiscale models is followed.

\begin{figure}[htbp]
\centering
\tikzstyle{vertex}=[draw, minimum width=30pt, minimum height=30pt, align=center]
\tikzstyle{container} = [draw, rectangle, inner sep=1em, fill=black, opacity=0.3]
\tikzstyle{circles} = [ellipse, draw, minimum height=50pt, minimum width=30pt]
\begin{tikzpicture}[transform shape]
\node(sample_x) at (-.75, 0) {\begin{tabular}{cc}$\xi$\end{tabular}};
\node[vertex](M) at (1, 0) {\begin{tabular}{cc}$M$\end{tabular}};
\node[vertex](mu) at (1, -2) {\begin{tabular}{cc}$\mu$\end{tabular}};
\draw[->] (M.240) -- node [anchor=east] {$u^{t_M}$} (mu.120);   
\draw[->] (mu.60) -- node [anchor=west] {$v^{t_{\mu}}$} (M.300);  
\draw[->] (sample_x) -- node [] {} (M);
\node(sample_u) at (2.9, 0) {\begin{tabular}{cc}$u^{t_{end}}$\end{tabular}};
\draw[->] (M) -- node [] {} (sample_u);
\end{tikzpicture}
\caption{A multiscale model as a collection of two coupled single scale models $M$ and $\mu$, where $\xi$ is a vector of the model inputs parameters, $u^{t_M}$ is the response of macro model $M$, $v^{t_{\mu}}$ is the response of micro model $\mu$, and $u^{t_{end}}$ is the model response at the final time step.}\label{fig:MSM}
\end{figure}
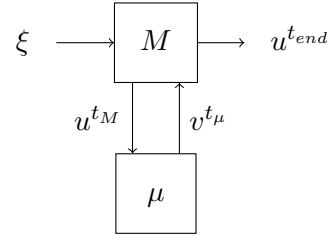

In Figure~\ref{fig:MSM}, an example of a multiscale model with two scales coupled via a scale bridging method is shown. The macro and micro models are denoted by the letters $M$ and $\mu$, respectively. The horizontal arrows are the model initialisation (left) and the final output (right), and the vertical arrows indicate the time execution loop, where at every iteration of the macroscale model the microscale model is executed until completion. 

In this work, one class of multiscale models is considered, where the macro and micro models have different time scales \cite{Chopard_Borgdorff_Hoekstra_2014}:
$$\Delta t_M \geq  n_{\mu} \Delta t_{\mu}; $$
$\Delta t_M$ and $\Delta t_{\mu}$ are the time steps at macro and micro levels, and $n_{\mu}$ is the number of timesteps at micro level with time step $\Delta t_{\mu}$, for each macro time step $\Delta t_M$. In this case, the models $M$ and $\mu$ can have overlapping or well separated spatial scales. In the case studies presented in section with examples, the case where $\Delta t_M = n_{\mu} \Delta t_{\mu}$ is considered, meaning that the micro and macro scale are touching each other on the scale separation map \cite{Hoekstra_2007}.

To initialise the multiscale model, the values of the model input parameters $\xi$ should be specified, where $\xi \in \R^n$ is a $n$-dimensional vector. At each macroscale simulation time point $t_M \in [0, t_{end}]$, the macro model calls the micro model, sending the initial state $u^{t_0}$, if $t_M=0$, or the result from the previous time step $u^{t_M - \Delta t_M}$, if $t_M>0$, with the Quantity of Interest (QoI) $u^{t} \in \R^m$ for $t \in [0,t_{end}]$. 
Next, the micro model is run with the time step $\Delta t_{\mu}$ until it reaches an equilibrium and it produces an output $v^{t_{\mu}}$ with $v^t \in \R^k$ for $t \in [0,t_{end}]$. 
Then, it sends this result back to the macro model, which produces an output $u^{t_M}$, and, then, the simulation time is increased by $\Delta t_M$. This process continues until the final simulation time $t_{end}$ is reached. Normally the microscale simulation is some fully resolved model that requires substantial computational resources. This microscale model is called over and over again at every time step of the macroscale model, rendering the microscale computations usually the most expensive part of a multiscale simulation.

Usually precise values of the inputs $\xi$ cannot be obtained. Hence, the model output $u^{t_M}$ contains an inherent uncertainty. Our goal is to estimate the output uncertainty accurately and in a minimal execution time. Since frequently an execution of multiscale models takes vast amount of time \cite{alowayyed2016multiscale}, straightforward black-box Monte Carlo methods can be prohibitive. Therefore, in the next section, a family of algorithms, which can perform multiscale UP in a more efficient way, will be introduced.



\section{Multiscale uncertainty propagation}\label{sec:UP_alg} 
As introduced above, the uncertain inputs are denoted by the vector $\xi$, and the output of interest is the response of the macro model $u^{t_M}$. The mean value ($ \mathbb{E}\left[ u^{t_M} \right] $) and the standard deviation ($\sigma \left[ u^{t_M} \right]$) need to be estimated as measures of uncertainty, assuming that the probability density $p_{u^t}$ is unimodal.

First, the estimation of the moments by a black-box Monte Carlo method is shortly described. Then, the semi-intrusive approach is introduced, and more specifically a semi-intrusive Monte Carlo method, and a metamodeling approach will be further explored.
 
\subsection{Plain Monte Carlo}\label{sec:MC}
 
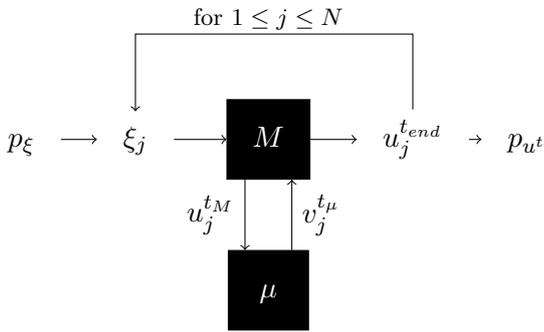
\begin{figure}[htbp]
\centering
\tikzstyle{vertex}=[draw, minimum width=30pt, minimum height=30pt, align=center, fill=black, text=white]
\tikzstyle{container} = [draw, rectangle, inner sep=1em, fill=black, opacity=0.3]
\tikzstyle{circles} = [ellipse, draw, minimum height=50pt, minimum width=30pt]
\begin{tikzpicture}[scale=1.0]
\node(u_init) at (-2.25, 0) {\begin{tabular}{cc}$p_{\xi}$\end{tabular}};
\node(sample_x) at (-.75, 0) {\begin{tabular}{cc}$\xi_j$\end{tabular}};
\draw[->] (u_init) -- node [] {} (sample_x);
\node[vertex](M) at (1, 0) {\begin{tabular}{cc}$M$\end{tabular}};
\node[vertex](mu) at (1, -2) {\begin{tabular}{cc}$\mu$\end{tabular}};
\draw[->] (M.240) -- node [anchor=east] {$u_j^{t_M}$} (mu.120);   
\draw[->] (mu.60) -- node [anchor=west] {$v_j^{t_{\mu}}$} (M.300);  
\draw[->] (sample_x) -- node [] {} (M);
\node(sample_u) at (2.9, 0) {\begin{tabular}{cc}$u_j^{t_{end}}$\end{tabular}};
\draw[->] (M) -- node [] {} (sample_u);
\path [draw, ->] (sample_u.north) -- ++(0,1cm) -- ++(-3.65cm, 0) -- (sample_x.north);
\node (for_loop) at (1, 1.6) {\footnotesize for $1 \leq j \leq N$};
\node(u_end) at (4.4, 0) {\begin{tabular}{cc}$p_{u^t}$\end{tabular}};
\draw[->] (sample_u) -- node [] {} (u_end);
\end{tikzpicture}
\caption{Black-box Monte Carlo method}\label{fig:qmc}
\end{figure}

An example of uncertainty estimation in the response of a multiscale model by a black-box Monte Carlo (MC) is shown in Fig.~\ref{fig:qmc}. We generate $N$ samples of uncertain inputs $\xi$ according to their probability distribution functions $p_{\xi}$, and run the model $N$ times with these inputs values. The model output is collected, and the $m$th moment of this output at the simulation time $t_M$ is estimated as
\begin{align}
\begin{split}
 \label{eq:UP_qmc_macro}
 \mathbb{E} \left[ \left( u^{t_M} \right)^m  \right] \approx \frac{1}{N} \sum^N_{j=1} \left( u_j^{t_M} \right)^m,
\end{split} 
\end{align}
where $u_j^{t_M}$ is the value of the macro model output when the model inputs have the values $\mathbf{\xi}_j$.
 
The quality of the estimates for the mean value and standard deviation by the MC method is usually provided by confidence intervals, which can be estimated by bootstrap \cite{DiCiccio1992}.

\subsection{Semi-intrusive methods}

The \textit{semi-intrusive} methods for multiscale UP are a family of algorithms, which employ the structure of the multiscale models in order to perform an efficient UP, that is, estimating the uncertainties with the comparable quality as the black box MC method, but with a substantially reduced execution time. According to the MMSF, instead of considering the whole multiscale model as a black-box, the model can be seen as a collection of coupled single scale black-box systems. Thus, the \textit{semi-intrusiveness} of the methods boils down to a limited inspection of the multiscale model, which is only up to the level of single scale components and their coupling.

\subsubsection{Semi-intrusive Monte Carlo}\label{sec:SIMC}

The semi-intrusive Monte Carlo (SIMC) is a Monte Carlo method with a reduced number of samples of the expensive component of the multiscale model. The remaining samples are obtained by interpolation. Additionally, a cross-validation is applied to test whether the approximation of the results does not lead to a large error in the estimates of uncertainty.

\begin{figure}
\centering
\tikzstyle{vertex}=[draw, minimum width=30pt, minimum height=30pt, align=center, fill=black, text=white]
\tikzstyle{sample_node}=[draw, minimum height=10pt, minimum width=10pt]
\begin{tikzpicture}[transform shape]
\node(u_init) at (-2.5, 0) {\begin{tabular}{cc}$p_{\xi}$\end{tabular}};
\node(sample_x) at (-1., 0) {\begin{tabular}{cc}$\xi_j$\end{tabular}};
\draw[->] (u_init) -- node [] {} (sample_x);
\node[vertex](M) at (1, 0) {\begin{tabular}{cc}$M$\end{tabular}};
\draw[->] (sample_x) -- node [] {} (M);
\node(sample_u) at (3, 0) {\begin{tabular}{cc}$u^{t_M}_j$\end{tabular}};
\draw[->] (M) -- node [] {} (sample_u);
\path [draw, ->] (sample_u.north) -- ++(0,1cm) -- ++(-4.cm, 0) -- (sample_x.north);
\node (for_loop) at (1, 1.6) {\footnotesize for $1 \leq j \leq N$};
\node(send_to_mu) at (3, -4) {\begin{tabular}{cc}$\xi_j$\end{tabular}};
\draw[->] (sample_u) -- node [anchor=west] { \{$u^{t_M}_{i}$\}$^{N_{\mu}}_{i=1} $ } (send_to_mu);
\node[vertex](mu) at (1, -4) {\begin{tabular}{cc}$\mu$\end{tabular}};
\draw[->] (send_to_mu) -- node [anchor=east] {} (mu);
\node(sample_mu) at (-1, -4) {$v^{t_{\mu}}_{j}$};
\draw[->] (mu) -- node [anchor=east] {} (sample_mu);
\path [draw, ->] (sample_mu.south) -- ++(0,-1cm) -- ++(4.cm, 0) -- (send_to_mu.south);
\node (for_loop2) at (1, -5.7) {\footnotesize for $1 \leq j \leq N_{\mu}$};
\node(interp_test) at (-1, -2) {\begin{tabular}{cc}Interpolation \& Testing\end{tabular}};
\draw[->] (sample_mu) -- node [anchor=east] { \{$v^{t_{\mu}}_{i} $\}$^{N_{\mu}}_{i=1}$ } (interp_test);
\draw[->] (interp_test) -- node [anchor=east] {\begin{tabular}{cc} \{$v^{t_{\mu}}_{i} $\}$^{N_{\mu}}_{i=1}$ $\bigcup$ \\ \{$\tilde{v}^{t_{\mu}}_{i} $\}$^{N}_{i=N_{\mu}+1} $\end{tabular}} (sample_x);
\node(u_end) at (4.75, 0) {\begin{tabular}{cc}$p_{u_t}$\end{tabular}};
\draw[->] (sample_u) -- node [] {} (u_end);
\end{tikzpicture}
\caption{Semi-intrusive Monte Carlo method with a smaller number of samples of the expensive microscale model}
\label{fig:simc}
\end{figure}
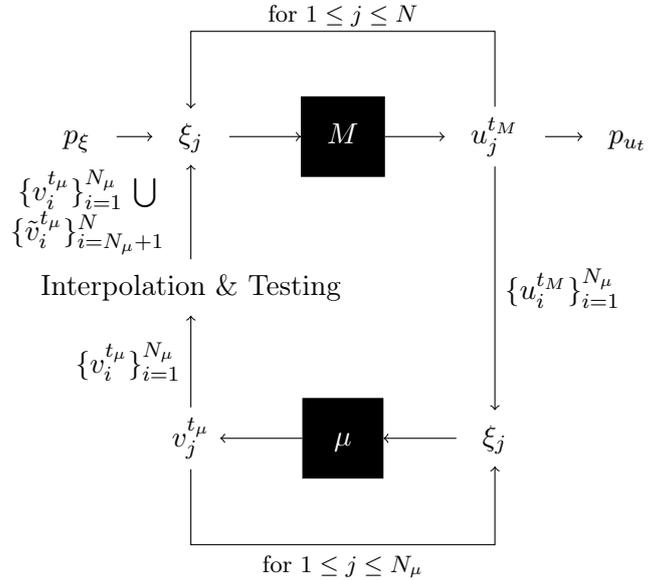

To perform the method the order of the time execution and the MC sampling is changed as in Fig.~\ref{fig:simc}. At each simulation time $t_M$ the macro model produces a sample of size $N$ of the QoI $\{u^{t_M}_{i}\}^{N}_{i=1}$. Then, it sends a set $\{u^{t_M}_{i}\}^{N_{\mu}}_{i=1}$ to the micro model, which in turn produces the set of outputs $\{v^{t_{\mu}}_{i}\}^{N_{\mu}}_{i=1}$. To obtain the samples $\{\tilde{v}^{t_{\mu}}_{i} \}^{N}_{i=N_{\mu}+1}$ an interpolation scheme is used. In this way, the expensive micro model is executed only $N_\mu$ times for $N_\mu \ll N$.

The moments of the QoI are approximated by
 \begin{align}
 \begin{split}
 \mathbb{E} \left[ \left( u^{t_M} \right)^m  \right] \approx \frac{1}{N} \left( \sum^{N_{\mu}}_{j=1} \left( u_j^{t_M} \right)^m
 + \sum^{N}_{j=N_{\mu}+1} \left( \tilde{u}_j^{t_M} \right)^m \right),
\end{split} 
 \end{align}
where $\tilde{u}_j^{t_M}$ is the result of the macro model with input of the micro model as a result of the interpolation.

Usually the interpolation method produces results which are not exact to the micro model response, and an error in the uncertainty estimates arises. Thus, a cross-validation on the sample of size $N_{\mu}$ must be performed, in order to estimate the effect of the micro model approximation. This test allows to decide whether it is safe to apply the SIMC method, or if instead the MC method with $N_{\mu}$ samples should be used.

\subsubsection*{Interpolation test}

In the process of the cross-validation of the results of interpolation the goal is to obtain confidence that the estimation of uncertainty by the SIMC method is close to the uncertainty of the original model response. In other words, the errors
\begin{align}
\begin{split}
\label{eq:error}
\epsilon_{\mathbb{E}} = \left |\mathbb{E}\left[u^{t_M}\right] - \mathbb{E}\left[\tilde{u}^{t_M}\right]\right| \text{ and }
\epsilon_{\sigma} = \left |\sigma \left[u^{t_M}\right] - \sigma\left[\tilde{u}^{t_M}\right]\right|
\end{split} 
\end{align}
must be small. To approximate these errors, their upper bounds will be estimated:
\begin{align}
\begin{split}
\label{eq:err_bound}
\epsilon_{\mathbb{E}} &\leq \mathbb{E} \left [| u^{t_M} - \tilde{u}^{t_M} | \right ] \text{ (by triangle inequality)},\\
\epsilon_{\sigma} &\leq \sigma \left [\left| u^{t_M} - \tilde{u}^{t_M} \right| \right] \text{ (by Cauchy-Schwarz inequality)}.
\end{split} 
\end{align}
This allows to study the random variable $\left| u^{t_M} - \tilde{u}^{t_M} \right|$, which can be estimated using the $N_{\mu}$ samples.

Denote by $f_{1,..,n-1}(\xi_n)$ the interpolation function, which approximates the micro model output corresponding to the one with the model input values $\xi_n$ using the set of original micro model outputs $\left\{v^{t_{\mu}}_{i} \right\}^{n-1}_{i=1}$. Call the set  of the interpolation results of the micro model output $\left\{\tilde{v}^{t_{\mu}}_{i} \right\}^{N_{\mu}}_{i=1}$, such that 
\begin{align}
\begin{split}
\tilde{v}^{t_{\mu}}_1 &= f_{2, \cdots, N_{\mu}}(\xi_1), \\
\tilde{v}^{t_{\mu}}_2 &= f_{1,3, \cdots, N_{\mu}}(\xi_2), \\
&\cdots \\
\tilde{v}^{t_{\mu}}_{N_{\mu}} &= f_{1, \cdots, N_{\mu}-1}(\xi_{N_{\mu}}).
\label{eq:int_function}
\end{split} 
\end{align}
Then, the macro model is run with these interpolated results for the micro model, resulting in the set $\left\{\tilde{u}^{t_{M}}_{i} \right\}^{N_{\mu}}_{i=1}$. Using this set and the set of the original macro model outputs $\left\{u^{t_{M}}_{i} \right\}^{N_{\mu}}_{i=1}$, the mean and the standard deviation of $\left| u^{t_M} - \tilde{u}^{t_M} \right|$ from inequalities~\ref{eq:err_bound} can be approximated:
\begin{align}
\begin{split}
\mathbb{E} \left [| u^{t_M} - \tilde{u}^{t_M} | \right ] \approx & \frac{1}{N_{\mu}}\sum^{N_{\mu}}_{i=1}{ \left| u_i^{t_M} - \tilde{u}_i^{t_M} \right|}, \\
\sigma \left[| u^{t_M} - \tilde{u}^{t_M}| \right] \approx &
\Bigg(\frac{1}{N_{\mu}-1}\sum^{N_{\mu}}_{i=1} \bigg( \left| u_i^{t_M} - \tilde{u}_i^{t_M} \right| \\ &
- \frac{1}{N_{\mu}}\sum^{N_{\mu}}_{i=1}{ \left| u_i^{t_M} - \tilde{u}_i^{t_M} \right|}\bigg)^2 \Bigg)^{\frac{1}{2}}.
\label{eq:err_bound_estimators}
\end{split} 
\end{align}

\begin{figure}[htbp]
\centering
  \includegraphics[width=0.5\textwidth]{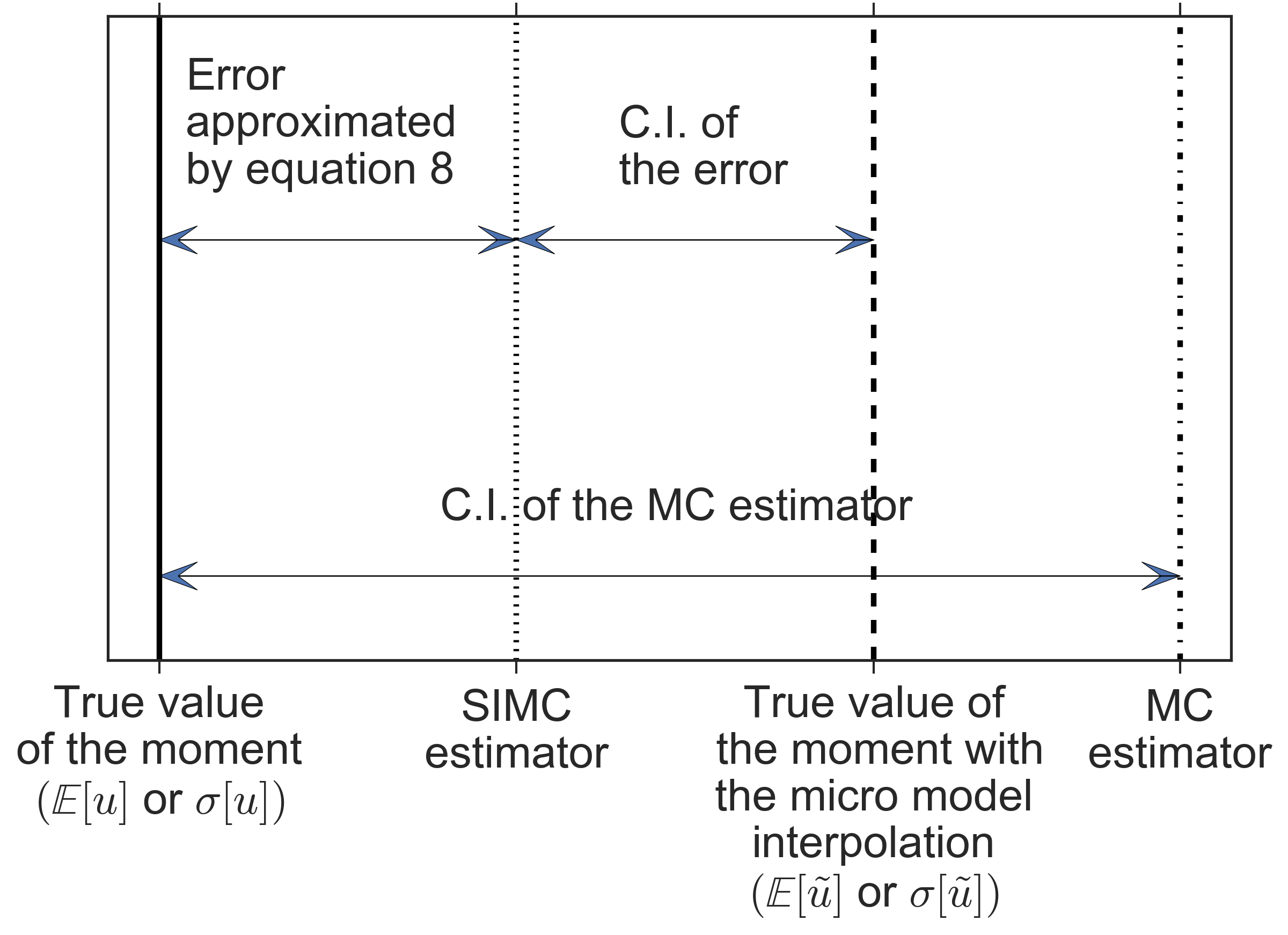}
  \caption{Comparison of the confidence in the estimators obtained by the SIMC and the MC with $N_{\mu}$ samples. Given a confidence level, the error of the SIMC method will not be larger than its approximation by Eq.~\ref{eq:err_bound_estimators} plus its confidence interval. In the illustrated case, when the upper endpoint of confidence interval of the SIMC error is lower than confidence interval of the MC estimates, the SIMC method is applied. Otherwise, the moments are estimated by the MC method.}
  \label{fig:SIMC_error}
\end{figure}

The idea of the interpolation test is to compare for each of the estimators the confidence interval of the MC result with $N_{\mu}$ samples and the error from Eq.~\ref{eq:err_bound_estimators} plus its confidence interval. In Figure~\ref{fig:SIMC_error}, an example, when the first is larger than second, is shown, and, in these cases, the results of the SIMC are accepted. Otherwise, another interpolation method can be tested, or the MC estimates from the $N_{\mu}$ samples are used.

\subsubsection{Metamodeling of a single scale model}\label{sec:Meta}

Surrogate modeling is a common approach to perform an efficient UP for computationally intensive systems at a reduced amount of time. The idea of these methods is to substitute the original system by its surrogate, which produces a similar output, but their computational time is lower. In the semi-intrusive multiscale metamodeling method, these techniques are applied to a single scale component, which takes the largest portion of the computational time \cite{Nikishova_2018}. In this way, the error introduced by the approximation is expected to be small when estimating the uncertainties of the multiscale model.

\begin{figure}[htbp]
\centering
\tikzstyle{vertex}=[draw, minimum width=30pt, minimum height=30pt, align=center]
\tikzstyle{container} = [draw, rectangle, inner sep=1em, fill=black, opacity=0.3]
\tikzstyle{circles} = [ellipse, draw, minimum height=50pt, minimum width=30pt]
\begin{tikzpicture}[transform shape]
\node(u_init) at (-2.25, 0) {\begin{tabular}{cc}$p_{\xi}$\end{tabular}};
\node(sample_x) at (-.75, 0) {\begin{tabular}{cc}$\xi_j$\end{tabular}};
\draw[->] (u_init) -- node [] {} (sample_x);
\node[vertex, fill=black, text=white](M) at (1, 0) {\begin{tabular}{cc}$M$\end{tabular}};
\node[vertex](mu) at (1, -2.) {\begin{tabular}{cc}$\tilde{\mu}$\end{tabular}};
\draw[->] (M.240) -- node [anchor=east] {$\tilde{u}_j^{t_M}$} (mu.120);   
\draw[->] (mu.60) -- node [anchor=west] {$\tilde{v}_j^{t_{\mu}}$} (M.300);   
\draw[->] (sample_x) -- node [] {} (M);
\node(sample_u) at (2.9, 0) {\begin{tabular}{cc}$\tilde{u}_j^{t_{end}}$\end{tabular}};
\draw[->] (M) -- node [] {} (sample_u);
\path [draw, ->] (sample_u.north) -- ++(0,1cm) -- ++(-3.65cm, 0) -- (sample_x.north);
\node (for_loop) at (1, 1.6) {\footnotesize for $1 \leq j \leq N$};
\node(u_end) at (4.4, 0) {\begin{tabular}{cc}$p_{u_t}$\end{tabular}};
\draw[->] (sample_u) -- node [] {} (u_end);
\end{tikzpicture}
\caption{Semi-intrusive multiscale metamodeling uncertainty propagation}\label{fig:meta}
\end{figure}
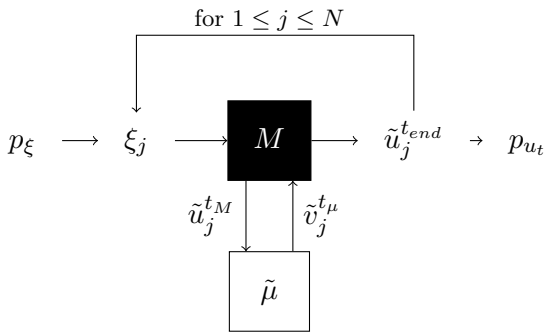

In Figure~\ref{fig:meta}, an example, where the micro model is substituted by a surrogate $\tilde{\mu}$, is shown. The rest of the multiscale model has the original form. However, since the micro model produces an approximate result $\tilde{v}_j^{t_{\mu}}$, the output of the macro model ($\tilde{u}_j^{t_M}$) is not the same as with the original model as well.

In this method, the error will always depend on the details of the model. It depends on the properties of the micro model, for example, smoothness, which determines how difficult it will be to approximate the original single scale model. Additionally, the error in the estimates of uncertainty also depends on how sensitive the result of the macro model is to the output of the micro model which is replaced by a surrogate. If, for instance, this sensitivity is low, it is reasonable to expect that the error introduced by the approximation is small. Of course, the error also depends on the method with which the surrogate is build. Next, several ways to obtain a metamodel $\tilde{\mu}$ of a single scale component are discussed.

A simplified physical metamodel is one of the options, where one seeks an approximate, maybe lower dimensional solver of the original problem, which would produce the result in a lower amount of time. For instance, this can be done by simplifying the physical description of the modelled process, or by solving the problem on a coarser computational mesh. The error produced due the approximation is problem specific, and one should perform analysis of these errors to preserve approximately correct estimates of uncertainties. 

The intrusive Polynomial Chaos (PC) is another method to build a surrogate. The approach is based on the analysis of the solver of the expensive single scale model, and the substitution by its stochastic representation \cite{DEB20016359,Wan_2005}. In general, this method works well and is easy to apply to linear and some non-linear problems. However, the solution can diverge when the method is applied, for example, to problems with a phase transitions \cite{pasini2013polynomial}. In such cases, one should perform a convergence analysis of the solution in order to obtain correct results of uncertainty estimation \cite{Xiu09fastnumerical}.

An inspection of the model solver is not required when a data-driven approach is applied. These methods are based on sampling the model output, and then on applying some regression method, such as for example the Gaussian processes \cite{Gorodetsky_2016,WANG2015159}, to obtain the model results in the rest of the function space. The regression methods have some limitations as well. For example, the Gaussian process regression works well only for smooth functions. However, one of the advantages of this method is that it provides an estimate on the precision of the prediction in the interpolated points, hence this allows to control the error of the approximated results.

\section{Examples}\label{sec:examples}

In this section, the results of uncertainty estimation for two systems based on reaction-diffusion dynamics are presented. The reaction-diffusion parameters are chosen such that characteristic times for the reactions are much short than for diffusion, rendering this a time-scale separated multiscale system. First, results obtained by the Monte Carlo (MC) method are presented, which are used as a reference solution. Next, the results obtained by a number of variants of the semi-intrusive methods are shown. The semi-intrusive multiscale UP methods that have been tested are SIMC with cubic interpolation, metamodeling with a data-driven surrogate build using the Gaussian Processes (GP) regression, and with a metamodel obtained by the intrusive Polynomial Chaos (PC), which is coupled to the non-intrusive PC. Additionally, results obtained by an intrusive Galerkin method are shown. In this way, the semi-intrusive methods are compared with both intrusive and non-intrusive techniques.

The mean value and the standard deviation of the concentration fields in the reaction diffusion systems were measured by these methods as uncertainty estimates. In the first case study, the MC results were obtained using $N=5000$ samples, which results in 95\% confidence interval for the standard deviation of at most 1.7\% of the estimator. In the second case study, the MC sample size is $N=30000$, which produces 95\% confidence interval for the standard deviation of at most 8\% of the estimator. Since the models outputs have nonnormal distributions, these confidence intervals were computed by bootstrap. 
For the rest of the methods, the parameters are indicated together with the presented results.

\subsection{Case study 1}
\label{sec:CS1} 
\begin{figure}[ht]
\centering
\begin{subfigure}[t]{0.49\textwidth}
\centering
  \includegraphics[width=\textwidth]{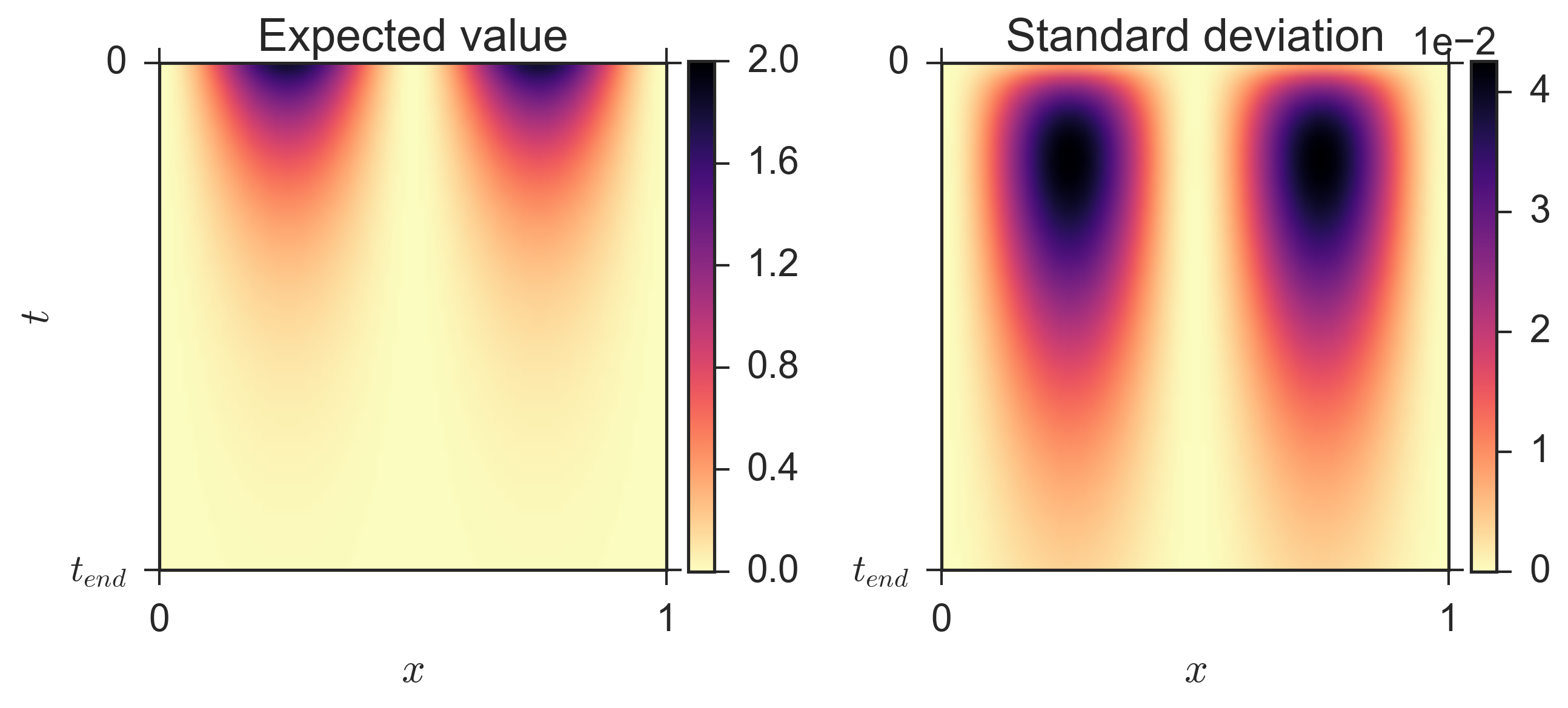}
  \caption{The multiscale system with $n_{\mu}=100$}
  \label{fig:UP_n1_100}
  \end{subfigure}
  \begin{subfigure}[t]{0.49\textwidth}
  \centering
    \includegraphics[width=\textwidth]{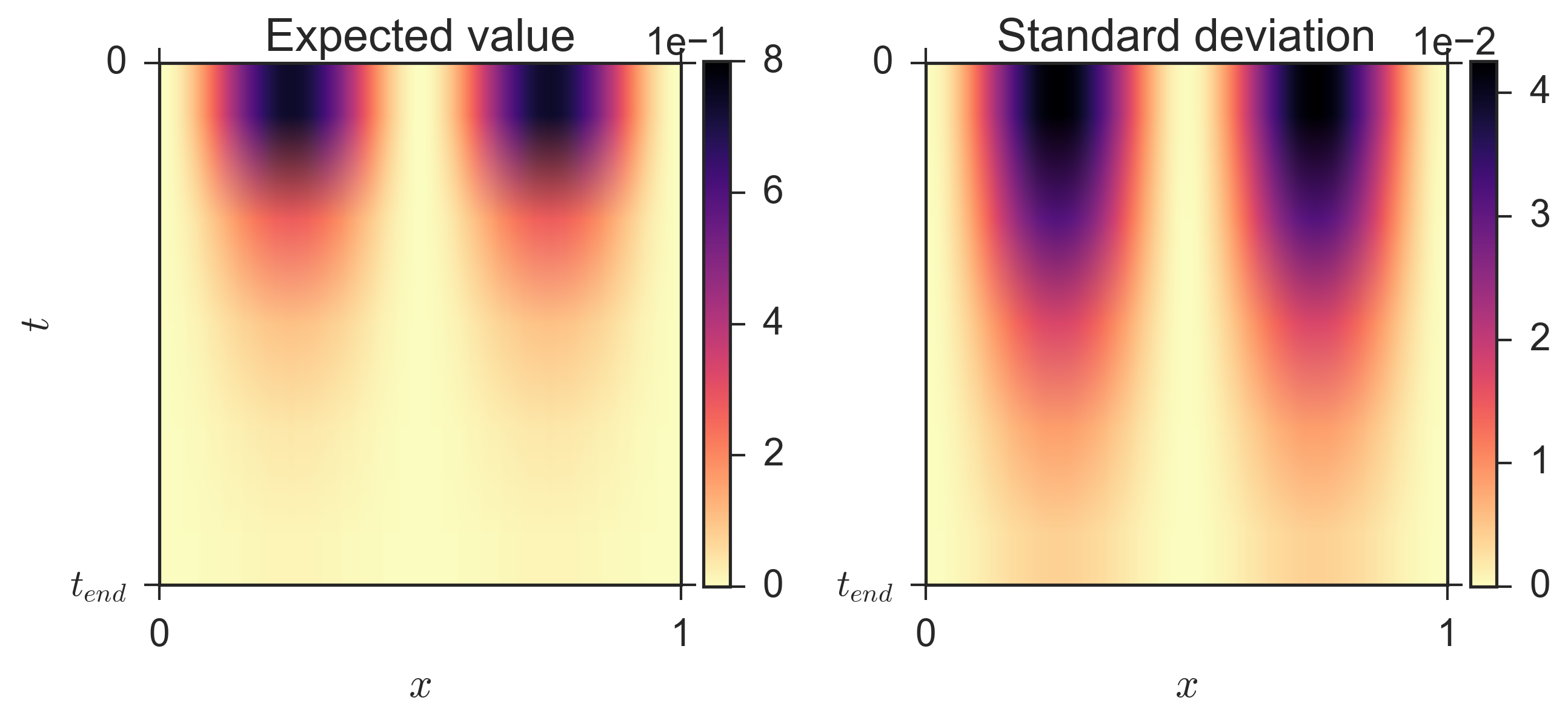}
  \caption{The multiscale system with $n_{\mu}=1000$}
  \label{fig:UP_n1_1000}
\end{subfigure}
\caption{The expected value and the standard deviation of the two systems estimated by the Monte Carlo method}
\label{fig:UP_ex_sol}
\end{figure}

The first case study is a 1D reaction-diffusion model with slow diffusion and fast reaction:
\begin{alignat*}{2}
&\frac {\partial u} {\partial t} =  d(\xi_1) \frac {\partial^2 u} {\partial x^2}+ &&k(\xi_2) u,
\end{alignat*}
for $x \in [0, 1]$ and $t \in [0, t_{end}]$, with
$$u(x,t=0, \xi_1, \xi_2) = \sin(\pi(4x -0.5)) +1, $$
$$u(x=0,t, \xi_1, \xi_2) = u(x=1,t, \xi_1, \xi_2), $$
where $d(\xi_1)$ and $k(\xi_2)$ are dimensionless diffusion and reaction coefficients with $10\%$ uncertainty. Uncertainty was estimated for results from model simulation with $n_{\mu} = \frac{\Delta t_{M}}{\Delta t_{\mu}} = 100$ and $n_{\mu} =1000$. The mean value of the diffusion coefficient is $\mathbb{E}[d(\xi_1)] = 4.05\cdot10^{-1}$, and the mean value of the uncertain microscale coefficient was set by $\mathbb{E}[k(\xi_2)] = \frac{n_{\mu} \mathbb{E}[d(\xi_1)]}{\Delta x^2}$. In the two experiments with different values of $n_{\mu}$ the space step is $\Delta x=10^{-2}$. The estimated mean and standard deviation of the response at different simulation time $t$ ($y$-axis) by the MC method is shown in Fig.~\ref{fig:UP_ex_sol}.

\begin{figure*}[h]
  \centering
  \begin{subfigure}[t]{0.47\textwidth}
        \includegraphics[width=\textwidth]{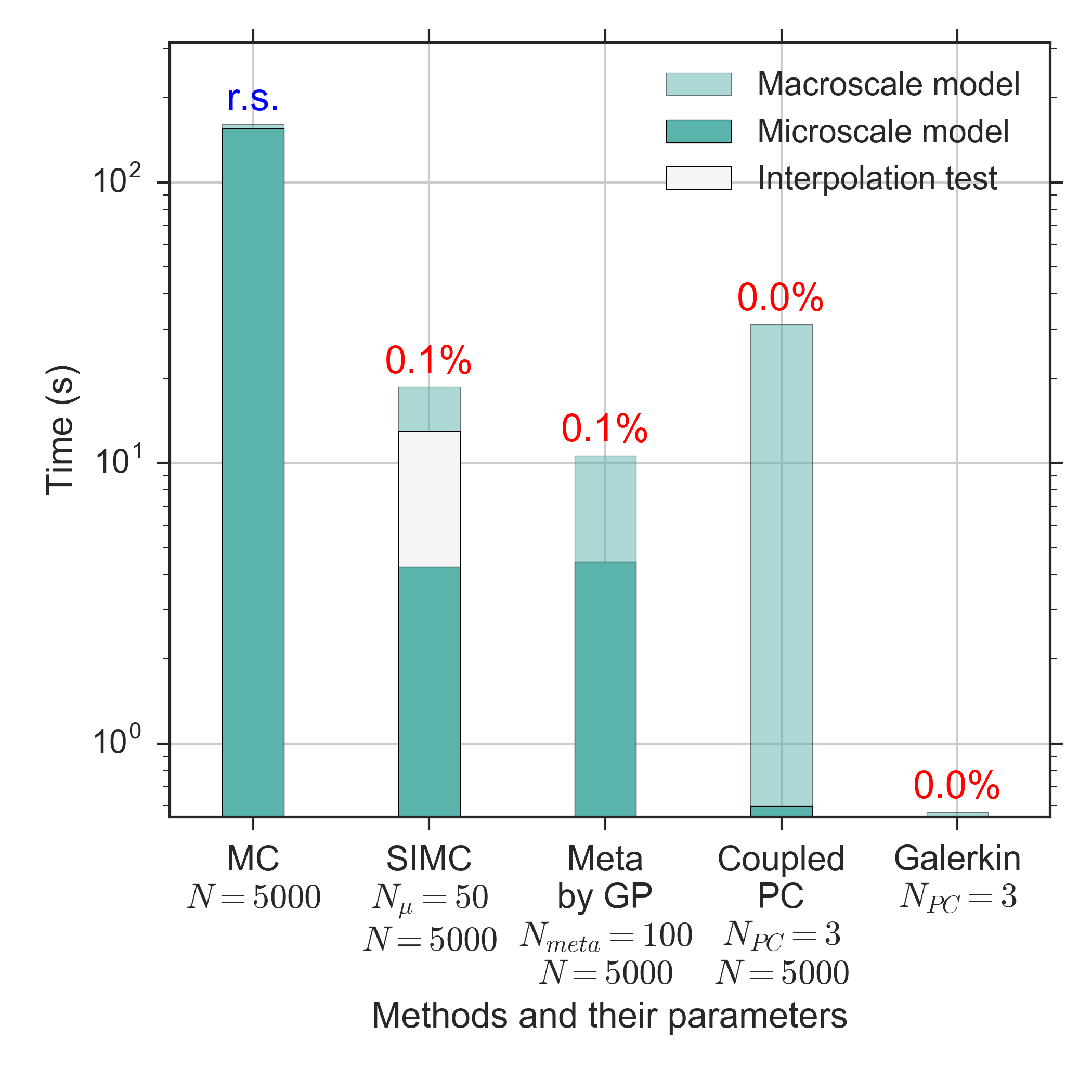}
        \caption{$n_{\mu}=100$}
        \label{fig:Comparison_n1_100}
   \end{subfigure}
   \begin{subfigure}[t]{0.47\textwidth}
  \centering
        \includegraphics[width=\textwidth]{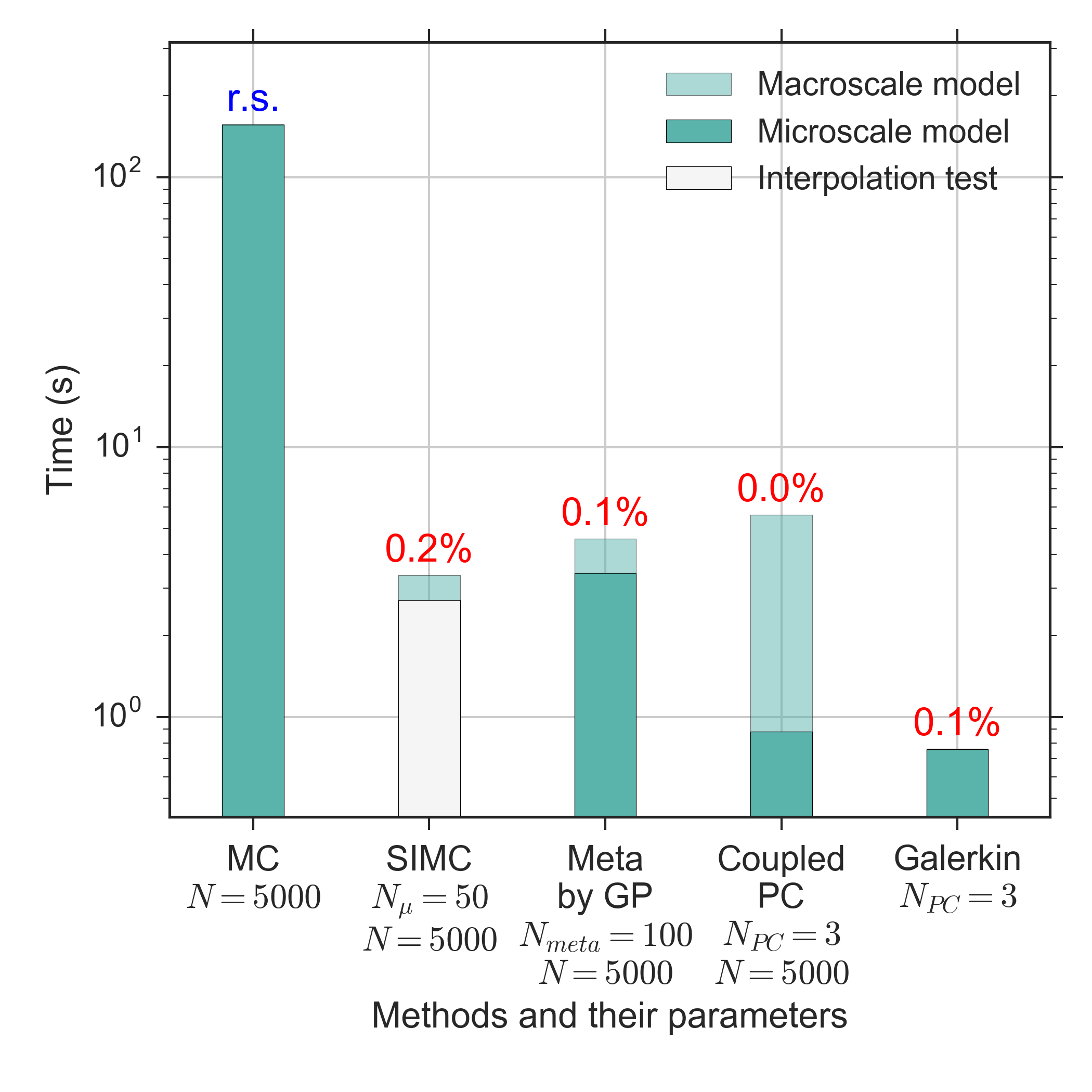}
        \caption{$n_{\mu}=1000$}
        \label{fig:Comparison_n1_1000}
   \end{subfigure}
   \caption{Comparison of UP methods in terms of execution time. Here and later, $N_{meta}$ is the number of samples used to build the data-driven metamodel, $N_{PC}$ is the truncated power in the Polynomial Chaos method, and "r.s" denotes the reference solution. The numbers above the bars are the mean relative errors in the results of the standard deviation obtained by the methods versus the MC results. The execution time broken down into time spent in respectively the macroscale model, the microscale model, and the interpolation test (only for the SIMC).}
   \label{fig:Comparison_case_1}
\end{figure*}

A performance comparison of the different UP approaches is presented in Fig.~\ref{fig:Comparison_case_1}. The methods are indicated on the $x$-axis, and the computational time in which the method produced the results is shown on the $y$-axis, which has a logarithmic scale. The semi-intrusive methods result in a speed up in comparison with the MC method, however, their time is still not so low as the computational time of the intrusive Galerkin method. Above each method bar the mean relative error in the estimate of the standard deviation from each of the methods relative to the MC method is indicated. The MC results serve as a reference solution (r.s.). In the results for both test systems the error does not exceed $0.2\%$.

In Table~\ref{table:time_case_1}, the computational times of the methods execution, portions of time spent on micro and macro components, and the speed up by the method versus the MC are included. The left part of the table contains data from the experiment with $n_{\mu}=100$, and the right part is with data for $n_{\mu}=1000$. In both cases, the MC uses most of the execution time on the micro model. The SIMC reduces this portion of time to about $70\%$ in the first case, and around $80\%$ in the second case. This leads that uncertainty for the first example were computed almost 9 times faster, and for the second almost 47 times faster than the MC method. This drastic reduction of computational time is because the interpolation takes a constant amount of time regardless the cost of the micro model execution. Therefore, if in the first example the interpolation time was relatively large, in the second example, this number is of little significance. Similarly, the metamodeling with GP and the coupled intrusive and non-intrusive PC compute about 15 times and 5 times faster, respectively, than the MC in the first example, and about 34 and 28 times faster in the second. Therefore, as higher the portion of the execution time of the micro model, as more efficiency is gained by the semi-intrusive methods. The Galerkin method is a highly efficient approach for this system, where it produces the result from 205 to 283 times faster than the MC.

\begin{table*}[ht]   
  \centering
      \caption{Computational time and speed-up in comparison with the MC method}\label{table:time_case_1}
      {
      \begin{tabular}{r||ccccc}
      \toprule
        \begin{tabular}{cc}Execution\\time\end{tabular}  & MC & SIMC & \begin{tabular}{cc}Meta-\\modeling\\by GP\end{tabular} & \begin{tabular}{cc}Coupled\\PC\end{tabular} & Galerkin\\
          \midrule 
           & \multicolumn{5}{c}{$n_{\mu}=100$} \\
          \midrule 
         $T^{total}$ (s)                 
         & 161.4  & 18.6  & 9.1   & 31.2  & 0.6   \\
         [0.3cm]
         $\dfrac{T^{\mu}}{T^{total}} 100 \%$ 
         & 96.4\%  & 69.7\% & 37.8\% & 1.9\%  & 96.0\% \\
         [0.3cm]
         $\dfrac{T^{M}}{T^{total}}  100 \%$   
         & 3.6\%   & 30.3\% & 62.2\% & 98.1\% & 4.0\%  \\   
         [0.3cm]
         $\dfrac{T^{total}_{MC}}{T^{total}_{method}}$ 
		 & 1.0   & 8.7   & 15.2  & 5.2   & 283.5 \\
          \midrule 
          & \multicolumn{5}{c}{$n_{\mu}=1000$} \\
          \midrule 
         $T^{total}$ (s)                 
         & 156.7  & 3.4   & 4.6   & 5.6   & 0.8   \\
         [0.3cm]
         $\dfrac{T^{\mu}}{T^{total}} 100 \%$ 
         & 99.5\%  & 80.7\% & 74.5\% & 15.8\% & 99.6\% \\
         [0.3cm]
         $\dfrac{T^{M}}{T^{total}}  100 \%$   
         & 0.5\%   & 19.3\% & 25.5\% & 84.2\% & 0.4\%  \\
         [0.3cm]
         $\dfrac{T^{total}_{MC}}{T^{total}_{method}}$ 
         & 1.0   & 46.8  & 34.3  & 28.0  & 205.5  \\
        \bottomrule
        \end{tabular}
        }
\end{table*}

\begin{figure*}
\centering
  \includegraphics[width=\textwidth]{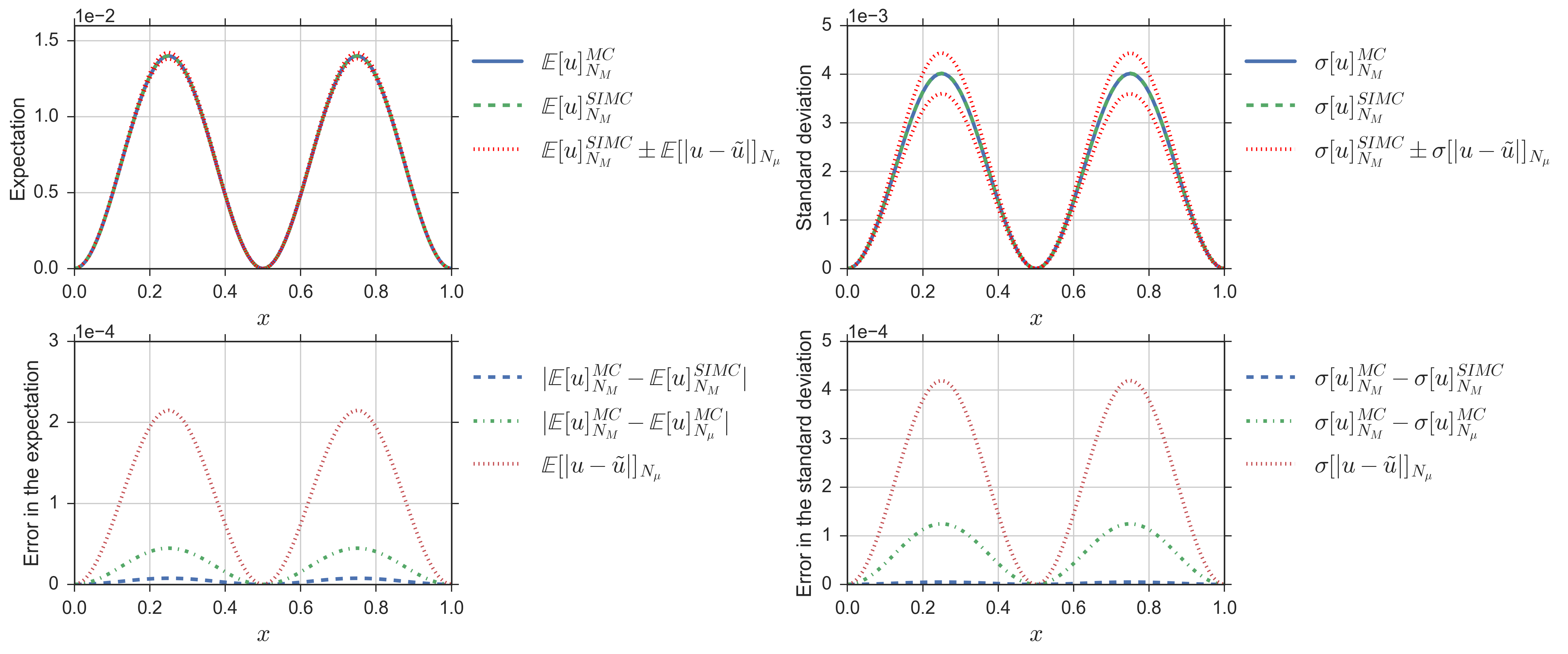}
  \label{fig:error_analysis_1000}
\caption{The error analysis in the expected value and the standard deviation at the final time step estimated by the SIMC method for the system with $n_{\mu}=1000$.}
\label{fig:SIMC_err_anal}
\end{figure*}

 
In Figure~\ref{fig:SIMC_err_anal}, an analysis of the error in the estimates of uncertainty at the final time step by the SIMC method is presented for the system with $n_{\mu}=1000$ (the system with $n_{\mu}=100$ shows a similar result). The upper plots show the estimates of the mean value (left) and standard deviation (right) by the MC and the SIMC methods, which show a good match for both estimates. Additionally, it includes the error estimates of the SIMC results, and, indeed, the MC estimates are within these bounds. The bottom plots show the errors in the estimates of uncertainty from the SIMC method and the MC method with $N_{\mu}$ samples in comparison with the MC method with $N$ samples. For both the mean and standard deviation, the SIMC results show lower error than the MC with $N_{\mu}$ samples. Indeed, in these examples, the interpolation test is passed.

 In this first case study, it is demonstrated that the proposed semi-intrusive methods are more efficient when applied to a multiscale model with a computationally expensive single scale component in comparison with the rest of the system. In the second example, the accuracy of the semi-intrusive UP methods is studied, where the methods are applied to a Gray-Scott model that has a highly non-trivial response.
\subsection{Case study 2}
\label{sec:CS2}

In this example, a two-dimensional Gray-Scott model \cite{Pearson1993} is studied:
\begin{align}
\begin{split}
\label{eq:Competition}
\frac{\partial u} {\partial t} &=  D_u  \nabla^2 u + F(\xi_1) (1 - u) -uv^2, 
\\
\frac{\partial v} {\partial t} &=  D_v \nabla^2 v -  (F(\xi_1)  + k (\xi_2))v + uv^2, \\
\end{split}
\end{align}
for space variables $x, y \subseteq[0,2.5]^2$, where $D_u$ and $D_v$ are dimensionless diffusion coefficients, $F(\xi_1)$ is a dimensionless feed rate, and $k(\xi_2)$ is the dimensionless rate constant of the second reaction. The reaction and diffusion processes are decoupled with $n_{\mu}=3$ with reaction faster than diffusion. The system has Neumann boundary conditions and initial conditions as follows \cite{Pearson1993}
\begin{align}
\begin{split}
v(t=0, x, y, \mathbf{\xi}) =
  \begin{cases}
   \frac{1}{4}\sin^2(4 \pi x)&\sin^2(4 \pi y),      \\ &  \text{if } x, y \subseteq[0.75,1.75]^2,\\
   0,  &  \text{otherwise,}\\
  \end{cases}
\\
u(t=0, x, y, \mathbf{\xi}) = 
  \begin{cases}
   -2v(t=0, x, & y, \mathbf{\xi}) + 1,       \\& \text{if } x, y \subseteq[0.75,1.75]^2,\\
   0,  & \text{otherwise.}\\
  \end{cases}
\end{split}
\end{align}

\begin{figure}[h]
\centering
\begin{subfigure}[t]{0.47\textwidth}
\centering
  \includegraphics[width=\textwidth]{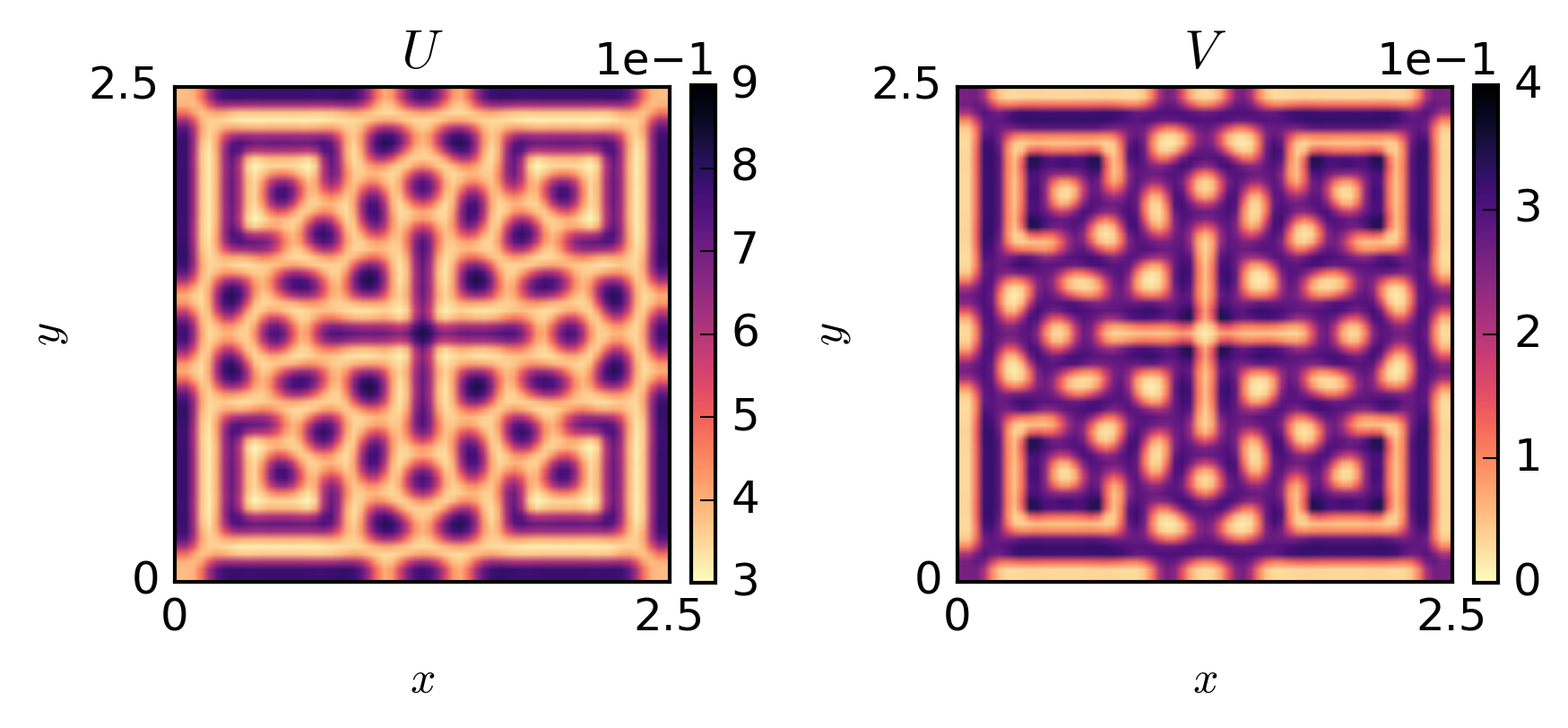}
  \caption{Output with $F=0.038885$ and $k=0.05148$}
  \end{subfigure}
\begin{subfigure}[t]{0.47\textwidth}
  \centering
    \includegraphics[width=\textwidth]{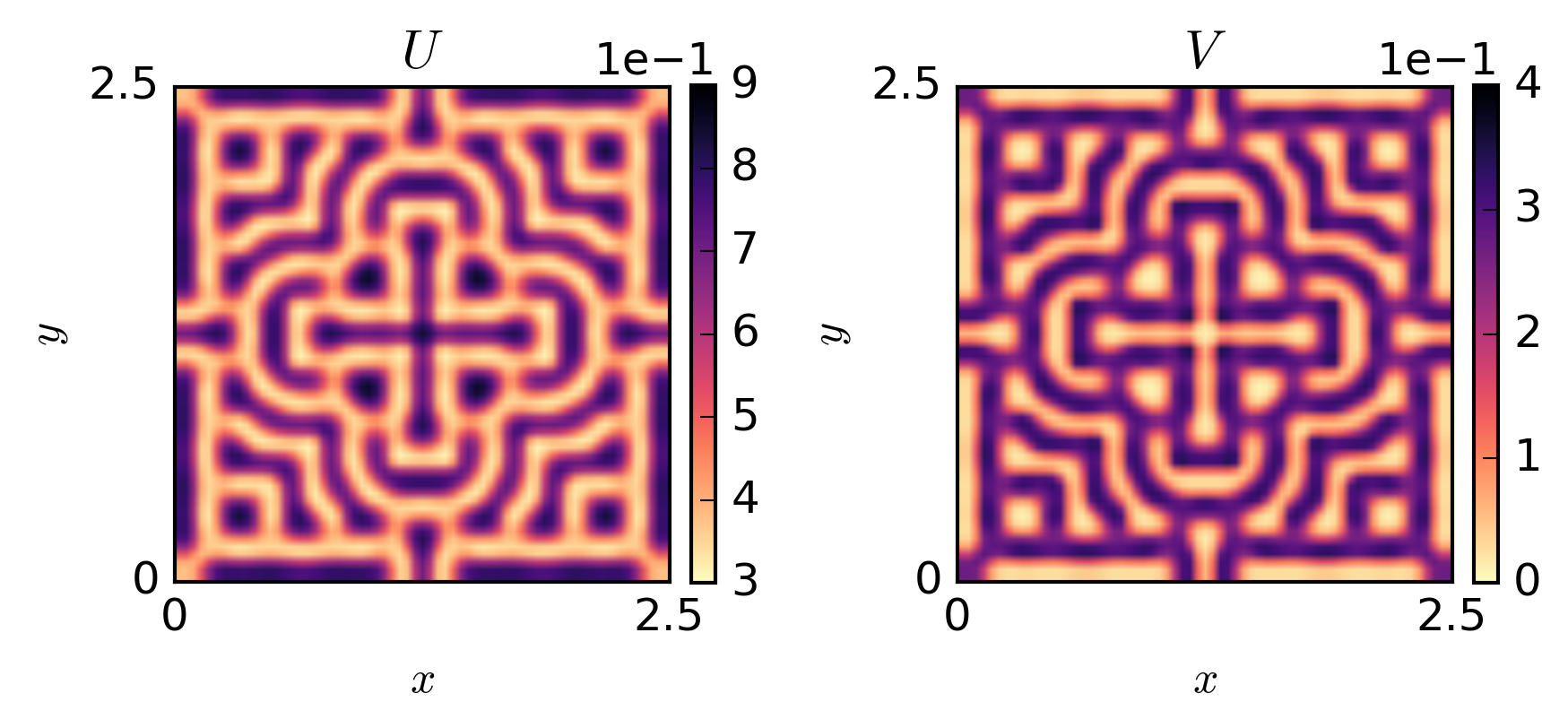}
  \caption{Output with $F=0.0385$ and $k=0.052$}
\end{subfigure}
\caption{Outputs of the Gray-Scott model with two different set of values for parameters $F$ and $k$.}
\label{fig:GS_two_res}
\end{figure}

The system is interesting to study, because the model output is very sensitive to the reaction coefficients $F(\xi_1)$ and $k(\xi_2)$. The model demonstrates a complex pattern formation with a transition map studied in \cite{Har-shemesh2016}. In Figures~\ref{fig:GS_two_res}, examples of two model outputs at the final simulation time with different sets of values of $F(\xi_1)$ and $k(\xi_2)$ are shown.

The model uncertain parameters are $F(\xi_1)$ and $k(\xi_2)$ with a uniform distribution and an $1\%$ variability range, and $D_u$ and $D_v$ are constants such that
\begin{align}
\begin{split}
\label{eq:inputs}
&\mathbb{E}[F(\xi_1)]=0.0385, \text{    }
\mathbb{E}[k(\xi_2)]=0.052, \text{    }\\ & D_u =2 \cdot10^{-5}, \text{    }  D_v=10^{-5}.
\end{split}
\end{align}

\begin{figure*}[h]
  \centering
  \includegraphics[width=0.8\linewidth]{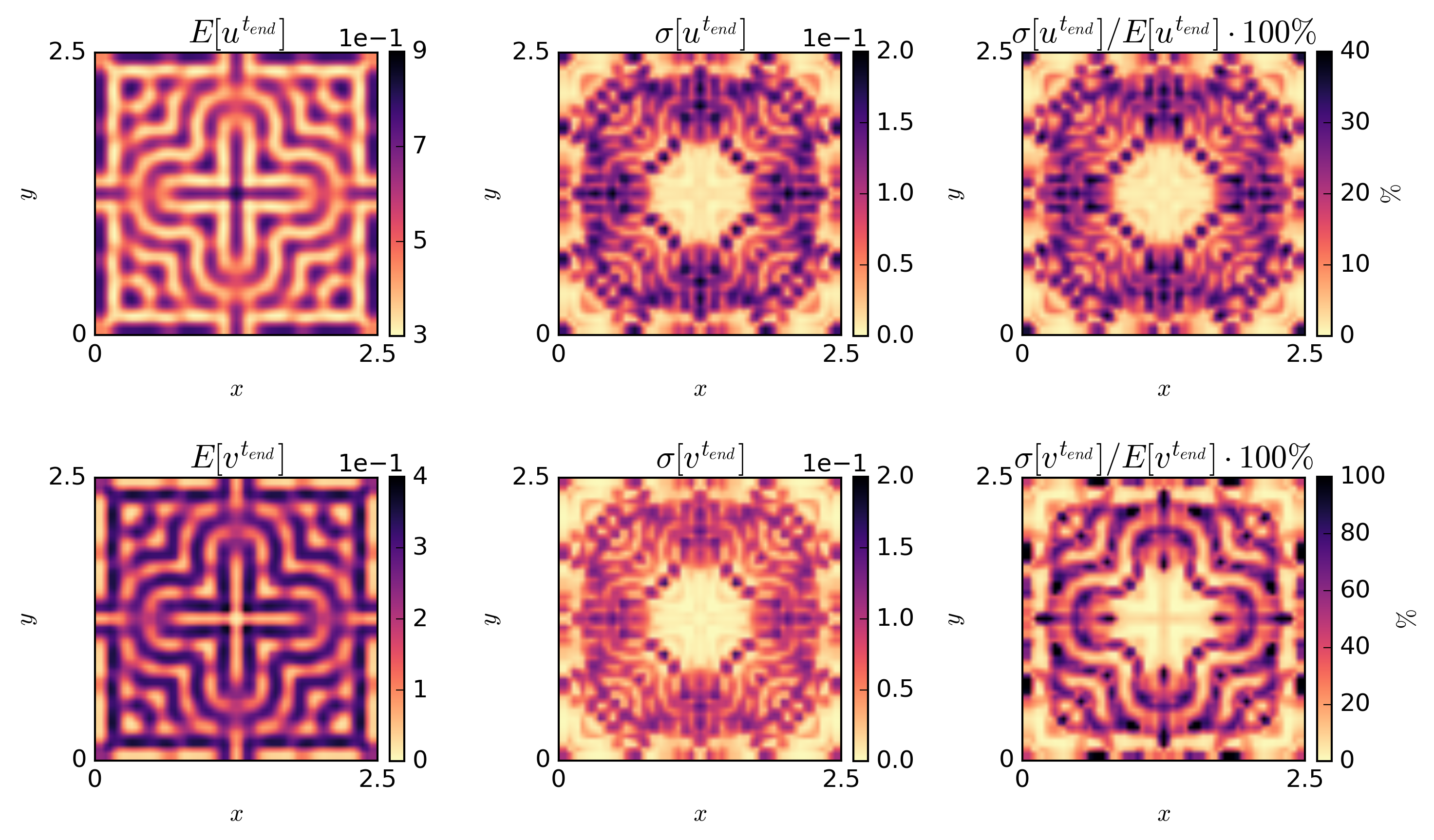}
  \caption{Uncertainty estimation result obtained by the MC method: the mean value (left column), the standard deviation (central column) and the coefficient of variation (right column) of the concentration $u$ (upper row) and $v$ (bottom row).}
  \label{fig:UP}
\end{figure*}

The UP result obtained with the MC of the final time step is presented in Fig.~\ref{fig:UP}. The results of the mean value are still quite close to the patterns from Fig.~\ref{fig:GS_two_res}, and this results for $u$ and $v$ are approximately reversed, i.e. $\mathbb{E}[u^{t_{end}}] \approx 1 - \mathbb{E}[v^{t_{end}}]$. At the same time, the standard deviations of $u$ and $v$ have a similar pattern. However, since the maximum value in space of $u$ is much greater than $v$, the relative uncertainty of $v$ represented by the coefficient of variation reaches $100\%$ at some locations, where this value for $u$ is about $36\%$.  

\begin{figure}[h]
  \centering
    \includegraphics[width=\linewidth,valign=t]{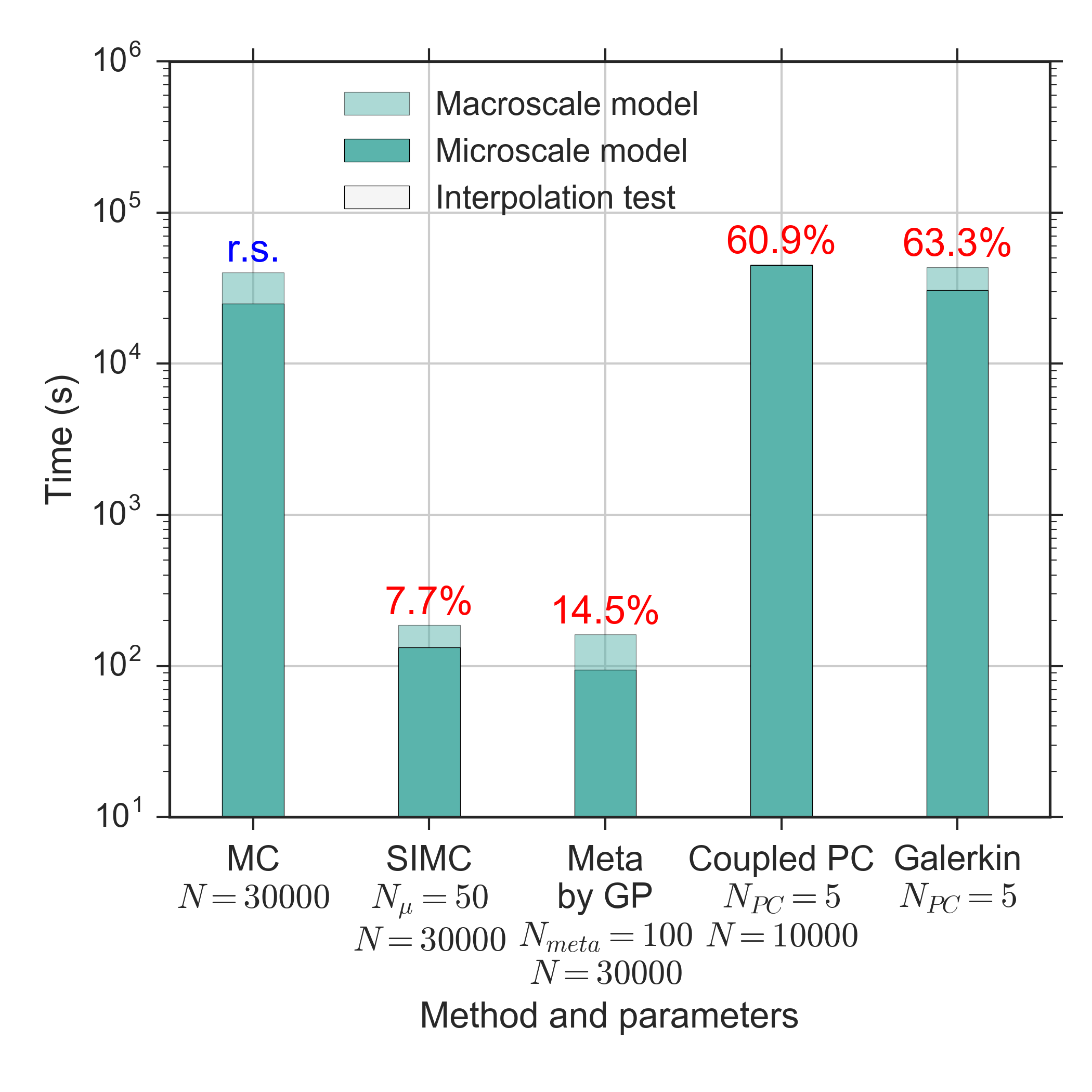}
    \caption{Comparison of the performance of the UP methods applied to the Gray-Scott model, where "r.s" denotes the reference solution, and the percentage indications above the columns are the mean relative error in the estimates of the standard deviation.}
    \label{fig:comparison}
\end{figure}

The comparison of the computational time and the error in the standard deviation by the UP methods are presented in Fig.~\ref{fig:comparison}, where the MC result is used as a reference solution. The SIMC and metamodeling with GP result in a significant drop in the execution time, and, in contrary, the coupled intrusive and non-intrusive PC and the Galerkin methods are more computationally expensive than the MC method. Moreover, the error in the results of the last two methods exceeds $60\%$. The high value of the error is due to nonlinear nature of the model, which cannot be approximated by a series of low order polynomials (in our case, the maximum order is $N_{PC} = 5$). The results obtained by the SIMC and the metamodeling with the GP are much closer to the MC results. In this example, the interpolation test in the SIMC is not passed, and, therefore, the $N_{\mu}=50$ samples are used to compute uncertainty with the MC, which produces $7.7\%$ error instead of $11\%$ when the result of the SIMC is accepted. Additionally, a more detailed error analysis of the SIMC method is given below.

In Table~\ref{table:time_comparison}, the total execution time of the UP methods and the portions of time taken by the micro and macro model executions are shown. The increase of the truncation degree in the PC and the Galerkin methods to decrease error makes the methods highly computationally expensive. In the SIMC and the metamodeling with GP, the portion of time spent on micro model is approximately the same, since, first, the system is not strictly multiscale, and, second, uncertainty was computed only for the last time step. Nevertheless, these two methods result in a significant reduction of the computational time, i.e. $216$ and $248$ times faster than the MC method, respectively.

\begin{table*}[h]
  \centering
  \caption{Computational time and speed-up in comparison with the MC method}\label{table:time_comparison}
  {
      \begin{tabular}{r|ccccc}
      \toprule
        \begin{tabular}{cc}Execution\\time\end{tabular}  & MC & SIMC & \begin{tabular}{cc}Meta-\\modeling\\by GP\end{tabular} & \begin{tabular}{cc}Coupled\\PC\end{tabular} & Galerkin\\
              \midrule 
   $T^{total}$ (s)                 
   & 40190.9 & 185.7 & 161.9 & 45341.5 & 43577.7 \\[0.3cm]
   $\dfrac{T^{\mu}}{T^{total}} 100 \%$                  
   & 62.0\%   & 71.6\% & 58.2\% & 99.05\%  & 70.6\%     \\[0.3cm]
   $\dfrac{T^{M}}{T^{total}}  100 \%$      
   & 38.0\%   & 28.4\% & 41.8\% & 0.95\%   & 29.4\%   \\ [0.3cm]
   $\dfrac{T^{total}_{MC}}{T^{total}_{method}}$  & 1.0     & 216.4 & 248.2 & 0.9     & 0.9  \\
  \bottomrule
  \end{tabular}
  }
\end{table*}

\begin{figure*}[h]
  \centering
    \includegraphics[width=\linewidth,valign=t]{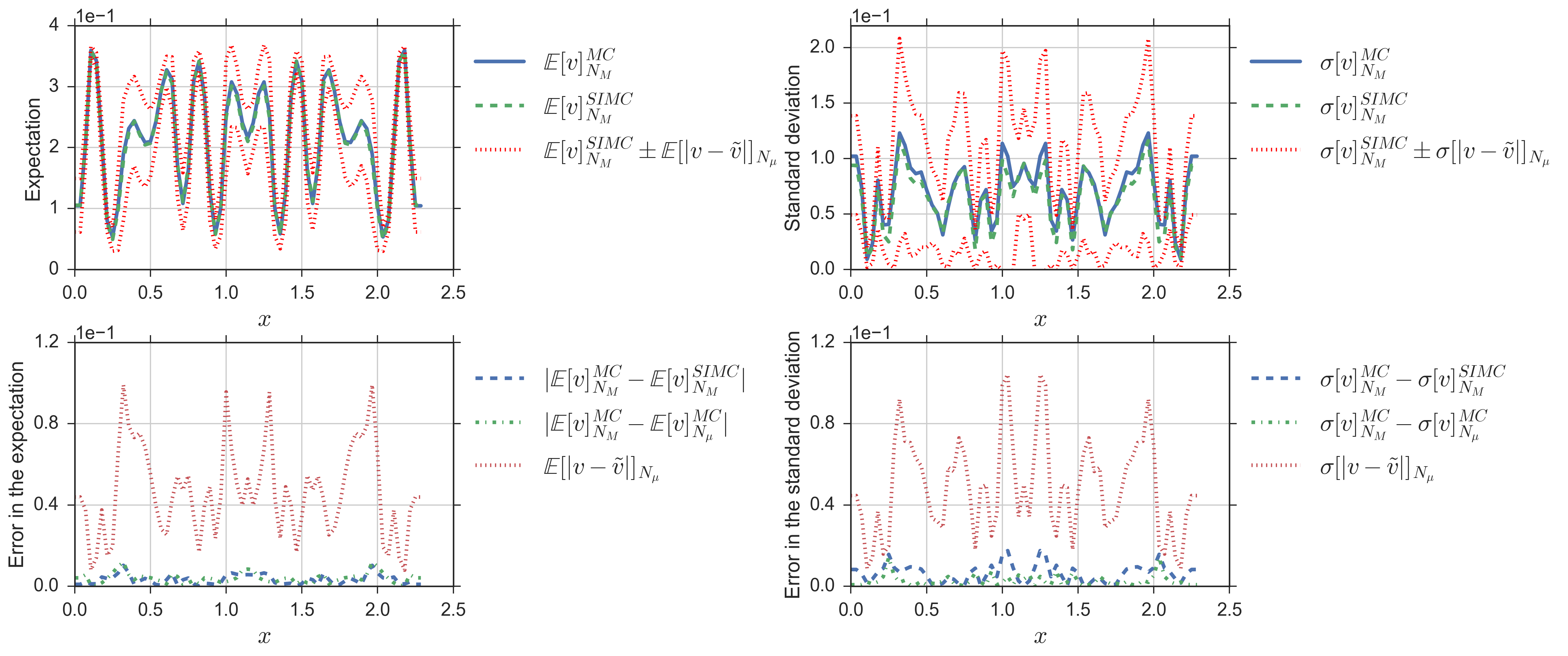}
    \caption{The SIMC error analysis for the concentration $v$ at the final time step and at $y=0.625$.}
    \label{fig:err_SIMC}
\end{figure*}

In Figure~\ref{fig:err_SIMC}, the results of the MC and SIMC methods are compared (top row), and the SIMC error is explored (second row) in the mean (left column) and the standard deviation (right column) for the concentration $v$ at the final time step and at $y=0.625$. The top left plot illustrates that the MC and SIMC estimates match well. However, the bottom left figure indicates that the error in the results of the $MC$ with $N_{\mu}$ samples is smaller that in the SIMC results at some locations. This explains why the interpolation test for the error in the mean value was not passed. The top right plot shows that the match with the MC results for the estimates of the standard deviation is worse, however, indeed the error is bounded by the standard deviation of the absolute difference (dotted line).
Moreover, in most of the locations the error in the results of the $MC$ with $N_{\mu}$ samples is smaller than in the estimates by the SIMC. According to the interpolation test, this MC result is accepted, therefore, the mean relative error in the standard deviation is $7.7\%$ instead of $11\%$. This shows that the interpolation test works. 

\section{Conclusions}
\label{sec:conclusions}
In this work, semi-intrusive multiscale strategies to perform an efficient uncertainty propagation (UP) for multiscale models are proposed and tested on two benchmark problems. It is shown that defining multiscale models according to MMSF, can help to reduce the computational time of the uncertainty study. The number of samples for the expensive single scale model can be reduced. This approach is called the semi-intrusive Monte Carlo (SIMC). Since one of the steps in this method is an interpolation test, the accuracy of estimates is controlled. Likewise, a metamodel of the microscale model can be built before performing UP, allowing to compute an approximate model response without running the expensive single scale model itself. Such surrogate model can be obtained by using a data-driven approach, or by substituting the original microscale model by its computationally cheap version. These methods can be applied to complex or unknown structure of single scale models. However, when one of the single scale models can be rewritten by its stochastic representation, the intrusive Galerkin method can be applied to this single scale, and the non-intrusive PC is applied to other components of the multiscale model.

In the first case study, the semi-intrusive approaches were tested on a one-dimensional reaction-diffusion model. It is shown that these methods help to reduce the computational time of UP while only inducing a small error in the uncertainty estimates. Moreover, two examples with different value of the number of micro time steps for each macro step ($n_{\mu}$) are studied. It is observed that for the larger value of $n_{\mu}$ the efficiency (in comparison to the non-intrusive MC) of the semi-intrusive methods is higher than for the smaller $n_{\mu}$. Thus, the strength of these methods is more visible, when the cost of the microscale model is much higher than the cost of the macroscale model, since the time of interpolation or to build a metamodel does not depend on $n_{\mu}$.

In the second case study, the semi-intrusive UP techniques are applied to a Gray-Scott model in order to test if the algorithms are effective and accurate for complex non-linear systems as well. It was observed that the coupled Galerkin method with non-intrusive Polynomial Chaos (PC) and the intrusive Galerkin methods converge very slowly and require a high degree of the truncated PC expansion. This makes the methods non-efficient in terms of computational time. The metamodeling approach with Gaussian Process regression instead decreased the computational time drastically, however, contained about $14\%$ error. The SIMC method produced a relatively small error ($7.7\%$), and significantly reduced the computational cost in comparison to the MC approach. Moreover, the reference solution usually is not available for real world problems, therefore, the magnitude of the estimates error can not be measured. The interpolation test tells whether the micro model can be approximated correctly with $N_{\mu}$ samples, and, the SIMC estimate accurately uncertainty using this interpolation function.

In this work, one generic case of multiscale models \cite{Chopard_Borgdorff_Hoekstra_2014} is considered, where the macro and microscale models have different time scales. However, the semi-intrusive multiscale methods can also be applied to other types of the multiscale models, including cases with spatial scale separation. In such cases one can expect that the computational needs for the microscale models can be much more substantial than the benchmarks in this paper, stipulating the absolute need of semi-intrusive methods to make multiscale UP a tractable problem.




\bibliographystyle{plain}
\bibliography{lib_UPMSM}{}

\begin{thebibliography}{10}
\providecommand{\url}[1]{{#1}}
\providecommand{\urlprefix}{URL }
\expandafter\ifx\csname urlstyle\endcsname\relax
  \providecommand{\doi}[1]{DOI~\discretionary{}{}{}#1}\else
  \providecommand{\doi}{DOI~\discretionary{}{}{}\begingroup
  \urlstyle{rm}\Url}\fi

\bibitem{alowayyed2016multiscale}
Alowayyed, S., Groen, D., Coveney, P.V., Hoekstra, A.G.: Multiscale computing
  in the exascale era.
\newblock Journal of Computational Science  (2017).
\newblock \doi{10.1016/j.jocs.2017.07.004}

\bibitem{borgdorff2014performance}
Borgdorff, J., Belgacem, M.B., Bona-Casas, C., Fazendeiro, L., Groen, D.,
  Hoenen, O., Mizeranschi, A., Suter, J., Coster, D., Coveney, P., et~al.:
  Performance of distributed multiscale simulations.
\newblock Phil. Trans. R. Soc. A \textbf{372}(2021), 20130407 (2014)

\bibitem{Borgdorff_2013}
Borgdorff, J., Falcone, J.L., Lorenz, E., Bona-Casas, C., Chopard, B.,
  Hoekstra, A.G.: Foundations of distributed multiscale computing:
  Formalization, specification, and analysis.
\newblock Journal of Parallel and Distributed Computing \textbf{73}(4),
  465--483 (2013).
\newblock \doi{10.1016/j.jpdc.2012.12.011}

\bibitem{BORGDORFF2014719}
Borgdorff, J., Mamonski, M., Bosak, B., Kurowski, K., Belgacem, M.B., Chopard,
  B., Groen, D., Coveney, P., Hoekstra, A.: Distributed multiscale computing
  with {MUSCLE} 2, the multiscale coupling library and environment.
\newblock Journal of Computational Science \textbf{5}(5), 719 -- 731 (2014).
\newblock \doi{https://doi.org/10.1016/j.jocs.2014.04.004}

\bibitem{Chopard_Borgdorff_Hoekstra_2014}
Chopard, B., Borgdorff, J., Hoekstra, A.G.: A framework for multi-scale
  modelling.
\newblock Philosophical Transactions of the Royal Society A: Mathematical,
  Physical and Engineering Sciences \textbf{372}(2021), 20130378–20130378
  (2014).
\newblock \doi{10.1098/rsta.2013.0378}

\bibitem{Chopard_2011}
Chopard, B., Falcone, J.L., Hoekstra, A.G., Borgdorff, J.: A framework for
  multiscale and multiscience modeling and numerical simulations.
\newblock In: Lecture Notes in Computer Science, pp. 2--8. Springer Berlin
  Heidelberg (2011).
\newblock \doi{10.1007/978-3-642-21341-0_2}

\bibitem{DEB20016359}
Deb, M.K., Babuska, I.M., Oden, J.: Solution of stochastic partial differential
  equations using {Galerkin} finite element techniques.
\newblock Computer Methods in Applied Mechanics and Engineering
  \textbf{190}(48), 6359 -- 6372 (2001).
\newblock \doi{https://doi.org/10.1016/S0045-7825(01)00237-7}.
\newblock
  \urlprefix\url{http://www.sciencedirect.com/science/article/pii/S0045782501002377}

\bibitem{DiCiccio1992}
DiCiccio, T.J., Martin, M.A., Young, G.A.: Analytical approximations for
  iterated bootstrap confidence intervals.
\newblock Statistics and Computing \textbf{2}(3), 161--171 (1992).
\newblock \doi{10.1007/BF01891208}

\bibitem{Weinan_2011}
E, W.: Principles of Multiscale Modeling.
\newblock Cambridge University Press (2011)

\bibitem{GERRITSMA20108333}
Gerritsma, M., van~der Steen, J.B., Vos, P., Karniadakis, G.: Time-dependent
  generalized polynomial chaos.
\newblock Journal of Computational Physics \textbf{229}(22), 8333 -- 8363
  (2010).
\newblock \doi{https://doi.org/10.1016/j.jcp.2010.07.020}.
\newblock
  \urlprefix\url{http://www.sciencedirect.com/science/article/pii/S0021999110004134}

\bibitem{Gorodetsky_2016}
Gorodetsky, A., Marzouk, Y.: Mercer kernels and integrated variance
  experimental design: Connections between {Gaussian} process regression and
  polynomial approximation.
\newblock SIAM/ASA Journal on Uncertainty Quantification (1), 796--828 (2016).
\newblock \doi{10.1137/15M1017119}

\bibitem{groen2014survey}
Groen, D., Zasada, S.J., Coveney, P.V.: Survey of multiscale and multiphysics
  applications and communities.
\newblock Computing in Science \& Engineering \textbf{16}(2), 34--43 (2014).
\newblock \doi{10.1109/MCSE.2013.47}

\bibitem{Har-shemesh2016}
Har-shemesh, O., Quax, R., Hoekstra, A.G., Sloot, P.M.A.: {Information
  geometric analysis of phase transitions in complex patterns: the case of the
  Gray-Scott reaction–diffusion model}.
\newblock Journal of Statistical Mechanics: Theory and Experiment p. 43301
  (2016).
\newblock \doi{10.1088/1742-5468/2016/4/043301}

\bibitem{hoekstra2014multiscale}
Hoekstra, A., Chopard, B., Coveney, P.: Multiscale modelling and simulation: a
  position paper.
\newblock Phil. Trans. R. Soc. A \textbf{372}(2021), 20130377 (2014).
\newblock \doi{10.1098/rsta.2013.0377}

\bibitem{Hoekstra_2007}
Hoekstra, A.G., Lorenz, E., Falcone, J.L., Chopard, B.: Towards a complex
  automata framework for multi-scale modeling: Formalism and the scale
  separation map.
\newblock In: Y.~Shi, G.D. van Albada, J.~Dongarra, P.M.A. Sloot (eds.)
  Computational Science -- ICCS 2007, pp. 922--930. Springer Berlin Heidelberg,
  Berlin, Heidelberg (2007)

\bibitem{Johnstone_2015}
Johnstone, R.H., Chang, E.T.Y., Bardenet, R., de~Boer, T.P., Gavaghan, D.J.,
  Pathmanathan, P., Clayton, R.H., Mirams, G.R.: Uncertainty and variability in
  models of the cardiac action potential: Can we build trustworthy models?
\newblock Journal of Molecular and Cellular Cardiology \textbf{96}, 49--62.
\newblock \doi{10.1016/j.yjmcc.2015.11.018}

\bibitem{karabasov2014multiscale}
Karabasov, S., Nerukh, D., Hoekstra, A., Chopard, B., Coveney, P.V.: Multiscale
  modelling: approaches and challenges.
\newblock Philosophical transactions. Series A, Mathematical, physical, and
  engineering sciences (2014).
\newblock \doi{10.1098/rsta.2013.0390}

\bibitem{Liu2016}
Liu, Y., Guo, J., Wang, Q., Huang, D.: {Prediction of Filamentous Sludge
  Bulking using a State-based Gaussian Processes Regression Model}.
\newblock Scientific Reports \textbf{6}, 31303 (2016).
\newblock \doi{http://dx.doi.org/10.1038/srep31303}

\bibitem{Le_Ma_tre_2010}
Ma{\^{\i}}tre, O.P.L., Knio, O.M.: Spectral Methods for Uncertainty
  Quantification.
\newblock Springer Netherlands (2010).
\newblock \doi{10.1007/978-90-481-3520-2}

\bibitem{Nikishova_2018}
Nikishova, A., Veen, L., Zun, P., Hoekstra, A.G.: Semi-intrusive multiscale
  metamodeling uncertainty quantification with application to a model of
  in-stent restenosis.
\newblock Philosophical Transactions A  (2018).
\newblock \doi{10.1098/rsta.2018.0154}

\bibitem{pasini2013polynomial}
Pasini, J.M., Sahai, T.: Polynomial chaos based uncertainty quantification in
  hamiltonian and chaotic systems.
\newblock In: 52nd IEEE Conference on Decision and Control, pp. 1113--1118
  (2013).
\newblock \doi{10.1109/CDC.2013.6760031}

\bibitem{Pearson1993}
Pearson, J.E.: {Complex patterns in a simple system.}
\newblock Science (New York, N.Y.) \textbf{261}(5118), 189--92 (1993).
\newblock \doi{10.1126/science.261.5118.189}

\bibitem{sloot2009multi}
Sloot, P.M., Hoekstra, A.G.: Multi-scale modelling in computational
  biomedicine.
\newblock Briefings in bioinformatics \textbf{11}(1), 142--152 (2009)

\bibitem{smith2013uncertainty}
Smith, R.C.: Uncertainty quantification: theory, implementation, and
  applications, vol.~12.
\newblock {SIAM} (2013)

\bibitem{Wan_2005}
Wan, X., Karniadakis, G.E.: An adaptive multi-element generalized polynomial
  chaos method for stochastic differential equations.
\newblock Journal of Computational Physics \textbf{209}(2), 617--642 (2005).
\newblock \doi{10.1016/j.jcp.2005.03.023}

\bibitem{WANG2015159}
Wang, B., Chen, T.: Gaussian process regression with multiple response
  variables.
\newblock Chemometrics and Intelligent Laboratory Systems
  \textbf{142}(Supplement C), 159 -- 165 (2015).
\newblock \doi{https://doi.org/10.1016/j.chemolab.2015.01.016}

\bibitem{Xiu09fastnumerical}
Xiu, D.: Fast numerical methods for stochastic computations: A review.
\newblock Communications in Computational Physics \textbf{5}(2-4), 242--272
  (2009)

\bibitem{Zhan2017}
Zhan, T., Fang, L., Xu, Y.: {Prediction of thermal boundary resistance by the
  machine learning method}.
\newblock Scientific Reports \textbf{7}(1), 7109 (2017).
\newblock \doi{10.1038/s41598-017-07150-7}

\end{thebibliography}

\end{document}